\documentstyle[prl,aps,multicol,epsf]{revtex}

\begin{document}
\draft
\title{Interaction of Mesoscopic Magnetic Textures \\
with Superconductors.}
\author{ Serkan Erdin$^{a}$,  Amin F. Kayali$^{a}$,\\
Igor F.Lyuksyutov$^{a,*}$ and Valery L. Pokrovsky $^{a,b}$}
\address{(a) Department of Physics, Texas A\&M University,\\
College Station, TX 77843-4242 \\
(b) Landau Institute for Theoretical Physics, Moscow, Russia}
\date{\today}
\maketitle

\begin{abstract}
\noindent 
We analyze magnetic textures in a thin magnetic film and screening currents 
induced by these textures in a superconducting film. The two films 
are supposed to be parallel and close each to other, but interact 
only via magnetic fields. We consider also vortices inside 
superconducting film and their interaction with magnetic textures.
As  examples of magnetic textures we consider magnetic dot and magnetic 
vortex. 
We derive the condition at which spontaneous formation of a vortex or a 
coupled 
vortex-magnetic defect is energetically favorable. It is proven that the normal
component of magnetic field generated by a magnetic dot changes sign in the 
presence of the superconducting vortex.
\end{abstract}

\begin{multicols}{2}

The fabrication and experimental study \cite{exp} of mesoscopic heterogeneous 
magnetic/superconducting systems together with recent theoretical predictions 
\cite{dotlp,vmlp,ln} open a new class of physical effects. Earlier two of us 
(I.F.L. and V.L.P.) proposed to separate superconductivity and magnetism in 
space employing the modern technique of nano-fabrication \cite{dotlp}. The 
proximity effect which suppresses both order parameters can be easily avoided 
by growing insulator oxide layers between ferromagnetic (FM) and 
superconducting (SC) components. Inhomogeneous magnetization of the magnetic 
texture generates magnetic field penetrating into the superconductor. The 
magnetic field from the SC currents interacts with magnetic subsystem 
\cite{vmlp,ln}. We have proposed different realizations of mesoscopic 
magneto-superconducting systems: arrays of magnetic dots on the top of a 
superconducting film \cite{dotlp}, magnetic/superconducting bi-layers 
\cite{vmlp}, magnetic nano-rods embedded into a superconductor \cite{ln}. 

In the majority of proposed systems a magnetic texture interacts with the 
superconducting current. An inhomogeneous magnetization generates magnetic 
field outside the magnet. The magnetic field generates screening currents in 
superconductors which, in turn, change the magnetic field. The problem 
must be solved self-consistently . In this article we develop a general 
formalism for interacting inhomogeneous magnetization (texture) and 
superconductors in the London's approximation. Employing this formalism, we 
find two elementary solutions for a circular magnetic dot on the top of a 
superconducting film, magnetized perpendicularly to it, and for a coupled 
magnetic and superconducting vortices. London's approximation works 
satisfactory since the sizes of all structures in the problem exceeds
 remarkably the coherence length $\xi$.

Recent theoretical works \cite{hwa,frey,bul} consider problems close to ours. 
The main difference between our work and the Ref. \cite{frey} is that they 
do not incorporate spontaneous vortices in the ground state. In comparison 
to the Refs. \cite{hwa,bul} our approach is more general and universal.
A more detailed comparison with these references will be given later.
 
So far only arrays of sub-micron 
size magnetic dots  covered by thin superconducting films have been 
prepared and studied experimentally \cite{exp}$^{**}$.
The effect of commencuracy on the transport properties was reported. This 
effect is not specific for 
magnets interacting with superconductors and was first found many 
years ago by Martinoli and his group \cite{martinoli}. 
Predicted effects specific for FM/SC systems, including topological 
instability of homogeneous state in bi-layers \cite{vmlp}, were not yet 
observed experimentally. 
\newline
\noindent
{\it General formalism}
The total energy of a stationary magnet-superconducting system reads:
\begin{equation}
E  = \int \bigl [\frac{{\bf B}^2}{8 \pi}  + 
\frac{m n_s {\bf v}_{s}^{2}}{2} - 
{\bf B}\cdot {\bf M} \bigr ]dV
\label{en}
\end{equation}
\noindent
where ${\bf B}$ is the magnetic induction,  ${\bf M}$  is the magnetization, 
$n_s$ is the density of superconducting carriers,  
${\bf v}_s$ is their velocity and $m$ is their effective mass. We 
assume that the superconducting density $n_s$ and the magnetization  
${\bf M}$ are separated in space. We assume also that the magnetic 
field  ${\bf B}$ and its vector-potential  ${\bf A}$  asymptotically 
turn to zero at infinity. Employing Maxwell equation 
${\bf \nabla}\times {\bf B} = 
\frac{4 \pi}{c} {\bf j}$ , and ${\bf B} = {\bf \nabla}\times {\bf A}$,  
the magnetic field energy can be transformed as follows:

\begin{equation}
\int \frac{{\bf B}^2}{8 \pi}  dV=
\int  \frac{{\bf j}\cdot  {\bf A}} {2 c}dV 
\label{en2}
\end{equation}
\noindent
The current  ${\bf j}$ can be represented as a sum:  
${\bf j}={\bf j}_s + {\bf j}_m$  
of the superconducting and magnetic currents, respectively:
\begin{equation}
 {\bf j}_s  = \frac{n_s\hbar e}{m}
\bigl ( \nabla \varphi -  \frac{2\pi}{\phi_0}{\bf A}\bigr );\hskip
.5truecm 
{\bf j}_{m} = c {\bf \nabla} \times {\bf M}.  
\label{curr}
\end{equation}
\noindent
In the gauge invariant Eq.\ref{curr} $\varphi$  is the phase 
of the SC carriers wave-function and  $\phi_0$  is the (SC) 
flux quantum. 
Plugging  Eq.(\ref{curr}) into  Eq.(\ref{en2}) and  Eq.(\ref{en}), 
after minor transformations, we arrive at following equation for the 
total energy:

\begin{equation}
H  = \int \bigl[\frac{n_s\hbar^2}{2m}({\nabla\varphi})^2 - 
\frac{ n_s\hbar e}{2mc}{\nabla\varphi} \cdot{\bf A} - 
\frac{{\bf B}\cdot {\bf M}}{2}\bigr] dV
\label{en3}
\end{equation}

The phase gradient ${\nabla\varphi}$ can be included into ${\bf A}$ as 
a gauge transformation with the exception of vortex lines, where ${\varphi}$  
is singular. Eq. (\ref{en3}) allows to separate the energy of vortices from
the energy of magnetization induced currents and fields and from 
the interaction energy.
\newline
\noindent
{\it Two dimensional textures and vortices}.  
Below we apply our general formalism to the case of two parallel films, one 
FM, another SC, both very thin and very close to each other. Neglecting their 
thickness, we assume that both films are located at $z = 0$. The super-carrier
density $n_s({\bf r})$ can be 
represented as  $n_s({\bf R}) = \delta (z)  {n}^{(2)}_s({\bf r})$ and 
the magnetization  ${\bf M}({\bf R})$  can be represented as 
${\bf M}({\bf R}) = {\delta(z)}{\bf m}({\bf r})$, 
where  ${n}^{(2)}_s({\bf r})$ is the super-carrier density per unit 
area, ${\bf m}({\bf r})$ is the magnetization per unit area and 
${\bf r}$ is the two-dimensional radius-vector. In what follows
${n}^{(2)}_s$  is assumed to be a constant and the index (2) is omitted.  
The energy (4) can be rewritten for this special case:

\begin{equation}
H  = \int \bigl[\frac{n_s\hbar^2}{2m}({\nabla\varphi})^2 - 
\frac{ n_s\hbar e}{2mc}{\nabla\varphi} \cdot{\bf a} - 
\frac{{\bf b}\cdot {\bf m}}{2}\bigr] d^2 {\bf r}
\label{en4}
\end{equation}
\noindent
where ${\bf a} = {\bf A}({\bf r},z = 0)$ and  
${\bf b} = {\bf B}({\bf r},z = 0)$. The vector-potential satisfies
Maxwell-London's equation:

\begin{eqnarray}
\nabla\times(\nabla\times {\bf A}) 
&=& - \frac{1}{\lambda} {\bf A} \delta (z) 
+ \frac{ 4\pi\hbar  n_se}{m c}{\nabla\varphi}\delta(z)\\ \nonumber
& +& 4\pi\nabla\times ({\bf m}\delta(z)) 
\label{vec}
\end{eqnarray}
\noindent
where $\lambda_{} = \lambda_L^2/d_S$, is the effective screening length 
for the SC film, $\lambda_L$ is the London penetration depth and 
$d_s$ is the SC film thickness. 

According to our arguments, the term proportional to  ${\nabla\varphi}$ in Eq.
(\ref{vec}) describes vortices.  A plane vortex characterized by its vorticity
$q$ and by position of its center on 
the plane ${\bf r}_0$ contributes a singular term 
$q\frac{\hat z\times ({\bf r}-{\bf r}_0) }{\vert {\bf r}-{\bf r}_0 \vert^2}
$ 
to  ${\nabla\varphi}$ and generates a standard vortex vector-potential: 
\begin{equation}
{\bf A}^{v}({\bf R}-{\bf r}_0) = 
\frac{q \phi_0}{2 \pi} 
\frac{\hat z\times ({\bf r}-{\bf r}_0) }{\vert {\bf r}-{\bf r}_0 \vert} 
\int_{0}^{\infty} 
\frac{J_1 ( k\vert{\bf r} - {\bf r}_0\vert) e^{-k| z |}}
{1 + 2 k\lambda}dk 
\label{vec2}
\end{equation}
Different vortices contribute independently into the vector-potential 
and magnetic field. Below we calculate the contribution of a
magnetic texture to the vector potential ${\bf A}^m ({\bf R})$ in a
special gauge $A_z = 0$. The magnetization vector field can be represented 
as follows:
\begin{equation}
{\bf m} ({\bf r}) = m_z ({\bf r}) \hat z + {\bf m}^\parallel ({\bf r}) +
{\bf m}^\perp ({\bf r})
\end{equation}
where  ${\bf m}^\parallel ({\bf r})$ is the gradient part of  ${\bf m} 
({\bf r})$ and  ${\bf m}^\perp ({\bf r})$ can be represented as a
vector product $\hat z \times {\bf \nabla} f$ ( $f$ is an arbitrary
function of ${\bf r}$ ). This representation is unique. The corresponding
contributions to ${\bf A}^m$ are :
\begin{equation}
{\bf A}^{m}({\bf R}) = {\bf A}^{\parallel} ({\bf r}) +  {\bf
A}^{\perp} ({\bf R})
\end{equation}
where 
\begin{eqnarray}
{\bf A}^{\parallel} ({\bf r}) &=& 2 \pi {\bf m}^\perp ({\bf r}) \times
{\hat z} \\
\nonumber
 A^{\perp}_{j}  ({\bf R}) &=& \int G_{j l} ({\bf R} - {\bf r}^\prime) m_l
( {\bf 
r}^\prime ) d^2 {\bf r}^\prime
\end{eqnarray}
where non-zero component $G_{l j} ({\bf R} - {\bf r}^\prime)$ are $({\bf
R} =
({\bf r}, z)
)$ : 
\begin{eqnarray}
G_{\alpha z} &=& \epsilon_{\alpha \beta} \frac{x_\beta -
x^{\prime}_\beta}{|{\bf r} - {\bf r}^\prime |} \frac{2}{\lambda^2} F_1 (
\frac{|{\bf r} - {\bf r}^\prime |}{\lambda} , \frac{z}{\lambda}) ; \hskip
0.2truecm \alpha = x, y  \\  \nonumber
G_{x y} &=& - G_{y x} = \frac{2}{\lambda^2}  F_0 (
\frac{|{\bf r} - {\bf r}^\prime |}{\lambda} , \frac{z}{\lambda})
\label{G}
\end{eqnarray}
Here : 
\begin{equation}
F_\gamma (\eta, \zeta) = \int_{0}^{\infty} \frac{J_\gamma ( \eta k) e^{- k
| \zeta | } k^2}{1 + 2 k} d k ; \hskip 0.5truecm \gamma = 0, 1
\label{F}  
\end{equation}
$\epsilon_{\alpha \beta}$ is the second rank antisymmetric tensor and 
$J_{\gamma}(x)$ are the Bessel functions.

Equations (\ref{G},\ref{F}) can be further simplified if the magnetization 
has only $z$
and $r$ components and both depend only on $r$. Then the vector-potential
${\bf A}^{m}_{\alpha} ( {\bf R} )$ is directed along azimuthal lines and
its only non-zero component reads : 
\begin{equation}
 A^{m}_{\phi} ({\bf R}) = \sum_{\alpha =0}^1 \int K_\alpha ({\bf R},{\bf
r}^\prime) m_\alpha
({\bf
r}^\prime) d^2 {\bf r}^\prime  
\label{vec6}
\end{equation}
where $\alpha =0$ stays for $z$, and  $\alpha =1$ stays for $r$,
\begin{equation}
K_{\alpha} ({\bf r},{\bf r}^\prime ) 
= \frac{4 \pi}{\lambda^2}  
f_{\alpha}(\frac{r}{\lambda},\frac{r^\prime}{\lambda},\frac{z}{\lambda})
\label{mag2}
\end{equation}
and
\begin{equation}
 f_{\alpha}({\xi},{\eta},{\zeta} )=
\int_{0}^{\infty}\frac{ J_{1} (\xi k)J_{\alpha}(\eta k) 
e^{-k|\zeta|} k^2}{1 + 2 k}dk.
\label{mag3}
\end{equation}

Santos {\it et al.} \cite{frey} have developed a formalism for calculation of 
magnetic fields and screening currents generated by magnetic textures in 
superconductors similar to ours. However, they did not consider singular 
current distributions, i.e. vortices. Bulaevsky {\it et al.} \cite{bul} 
considered a special system of prepared domains in a thick magnetic slab 
contacting with a semi-infinite superconductor and generating vortices in it. 
The spontaneous formation of domains and chains of vortices was discussed 
earlier \cite{spie,vmlp} for a bi-layer composed from very thin films.

\begin{figure}
\begin{center}
\epsfysize=1.3truein
\epsfbox{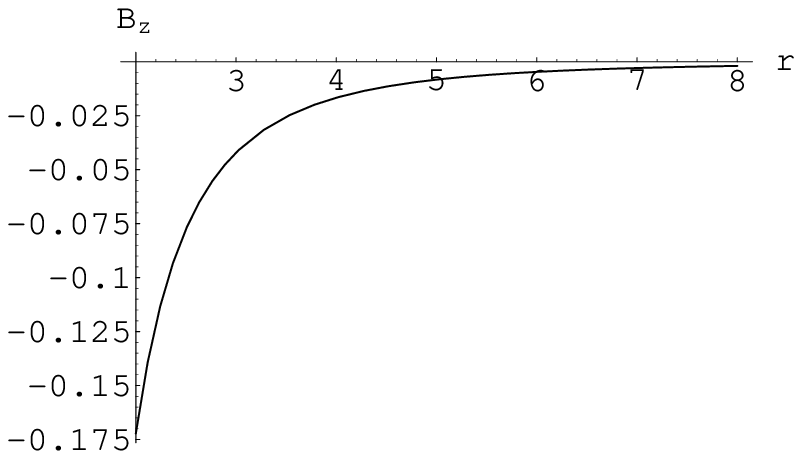}
\end{center}
\begin{center}
\vskip .2truecm
\epsfysize=1.3truein
\epsfbox{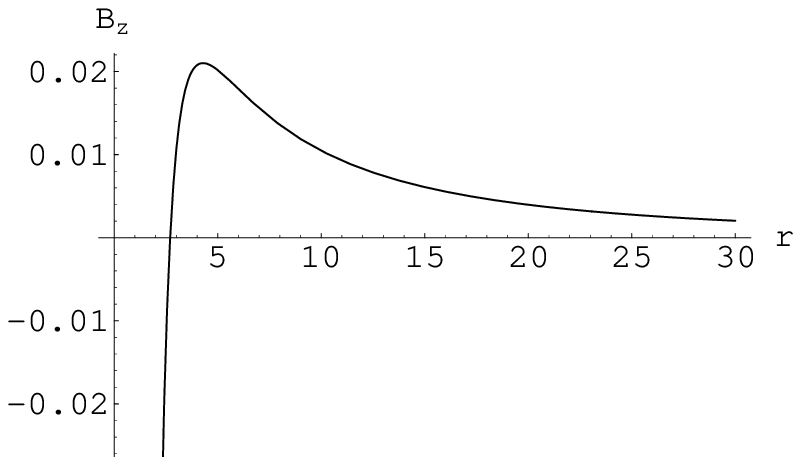}
\end{center}
\caption{Normal to the dot magnetization as a function of
distance from the  dot's center measured in the units of dots 
radius $R$. Top: without  vortex, for  $\lambda = 20R$;
Bottom: with vortex   
the case when  $q\phi_0/ (4 \pi^2 m ) =4$ 
and  $\lambda = 20R$.
\label{vor}}
\end{figure}

{\it Magnetic Dot}.
\noindent For infinitely thin circular  magnetic dot with magnetization
${\bf m} = m \hat z$ and radius $R$ the magnetic field can be 
calculated using Eq.\ref{vec6},\ref{mag2}. Its z-component has the form: 
\begin{eqnarray}
B_z (r,z) &=&  4 \pi\lambda m R \int_{0}^{\infty}  J_0 (kr) 
J_1 (kR) e^{- k |z|} \frac{k^2 dk}{1+2\lambda k} \nonumber \\
 &+& \frac{ q \phi_0}{2 \pi}
\int_{0}^{\infty} \frac{J_0 (rk) ke^{-k|z|}}{1 + 2 \lambda k} d k
\label{mf}
\end{eqnarray}
Numerical calculation based on Eq.\ref{mf} show that $B_z$ on the film $(z=0)$
changes sign at some $r > R$ if q is positive 
(see Fig.\ref{vor}), but it is negative everywhere at $r > R $ if 
$q=0$. The physical reason for this behavior is simple: the dot itself, 
without vortex currents, generates dipolar field which has the sign
opposite to dipolar moment in the plane passing through the dipole and 
perpendicular to it. The SC current resists this tendency. At small distances 
from the center the screening is neglegible and the field generated by dot 
dominates. At large distances $r\gg \lambda$ the field generated by a vortex 
decays like $1/r^3$, whereas the screened field of the dot decays as $1/r^4$.
The position of node $r_0$ can be easily determined if $R\ll\lambda$: then 
$r_0=2\pi\sqrt{mR^2\lambda /\Phi_0}$. Thus, the measurement of magnetic field 
near the film may serve as a diagnostic tool to detect SC vortex bound by the 
dot. The experimental measurements \cite{crabtree} demonstrated that $B_z$ 
changes sign, i.e. the dot indeed generates the vortex. 

In the presence of a vortex with charge $q$ , the energy of the system 
can be calculated using Eqs.(\ref{en4}, \ref{mf}). The result is:
\begin{equation}
E = q^2 \epsilon_0 ln \frac{\lambda}{\xi} - 
q\epsilon_m\frac{R}{2\lambda} - 
2 \pi m^{2} R ln \frac{R}{a}
\label{nvdot}
\end{equation}
where $\epsilon_m = m\Phi_0$ is the characteristic SC/FM interaction energy.
The vortex becomes energy favorable if $\epsilon_mR/(2\lambda)>\epsilon_0$
or $mR>[\Phi_0 /(8\pi^2)]\ln(\lambda /\xi)$.
The ratio 
\begin{equation}
\delta = \frac{\epsilon_m}{\epsilon_0} =
Sg\frac{2n_m d_m}{n_s d_s}  
\label{mdot}
\end{equation}
is the relative strength of the SC/FM coupling \cite{vmlp}. 
In Eq. (\ref{nvdot})  $n_m$ is the  density of magnetic atoms,
$d_{m,s}$ are the thicknesses of magnetic and superconduccting films, $S$ 
is the value of an elementary spin in the magnet and $g$ is the Lande factor. 
Far from the threshold the vorticity $q$ generated
by the magnetic dot is an integer closest to the value 
\begin{equation}
\tilde q = \frac{\delta}{4 \ln \frac{\lambda}{\xi}}
\frac{R}{\lambda} 
\label{nv}
\end{equation}
The logarithmic factor
in Eq. (\ref{nv}) varies in the range 3-8. For typical values  
$R = 10^{-4}$cm, $d_m = 10^{-6}$cm, and  $n_m = 10^{22}$cm$^{-3}$,
we find $\tilde{q} > 1$. Eq. (\ref{nv}) shows that, apart the logarithmic 
factor, the SV generation is controlled by properties of the 
magnetic film. The characteristic scale in superconducting film is $\lambda$. 
For this reason, at $R \approx\lambda$, the condition $\delta\approx 1$ gives 
an approximate criterion for the spontaneous vortex formation. 

Sasik and Hwa \cite{hwa} mimicked the magnetic dots by magnetic dipoles.
This approach does not reflect properly the real dot geometry. As it is 
seen from our results, the magnetic field and energy can not be expressed in 
terms of the total dot magnetic moment ($\pi mR^2$) only.

\noindent{\it Coupled Magnetic/Superconducting Vortices}.
Earlier we predicted spontaneous 
formation of the pairs of bound superconducting and magnetic 
vortices in the case of bi-layer consisting from superconducting 
and magnetic film with easy plane anisotropy \cite{vmlp}. A detailed
analysis given below confirms this prediction, but reveals more 
restrictive criterion for this phenomenon.
The exchange Hamiltonian of the  magnetic layer reads:
\begin{equation}
{\cal H}_M = \frac{J_M}{2}\int{d^2{\bf r}}({\nabla \bf n})^2
\label{h1} 
\end{equation}
where ${\bf n}({\bf r})$ is the unit 2D vector directed along the 
local magnetization. Further we assume that the vector ${\bf n}$ lays 
in plane (XY symmetry.) In a single magnetic vortex (MV) the local 
magnetization is directed along the radius: 
${\bf n}({\bf r}) = {\bf r}/r $. The energy (\ref{h1}) is invariant
under a global rotation of all spins by the same angle.
The MV energy  grows logarithmically with the system size $L:$ 
$E\approx \pi J_{M}\ln({L}/{\xi})$.
We employ general formulae, (\ref{vec}, \ref{mag2},\ref{mag3}) 
to calculate the vector-potential, magnetic field and the total energy. The 
vector-potential induced by the vortex, including the effect of 
screening currents is:
\begin{equation}
 A_{\varphi}({\bf R}) = -4\pi
m\lambda\int_{0}^{\infty}\frac{J_1(k r)}
{1+2\lambda k}e^{-k|z|}dk
\end{equation}
Employing the standard vortex vector potential Eq. \ref{vec}
with $R=0$ and eq. (\ref{en4}), we calculate the energy of the MV-SV 
pair with logarithmic accuracy: 
\begin{equation}
E = q^2\epsilon_0\ln(\frac{\lambda}{\xi})-
q\frac{\epsilon_m}{2}\ln(\frac{L}{\lambda})+ 8\pi^2 m^2 \lambda
\ln(\frac{L}{\lambda})
\label{nv2}
\end{equation}
The exchange energy of the magnetic vortex $E_{ex}$ is typically 
by 1-2 order of magnitude smaller than $\epsilon_0$ and can be
neglected. The energy is minimal at $q$ being an integer part of the value 
$\tilde{q}=\frac{m\phi_0\ln(\frac{L}{\lambda})}{2\epsilon_0 \ln(\frac{\lambda}
{\xi})}$.
The coupled pair of SC/FM vortices becomes energy favorable when minimum is 
realized at $q=1$. It means that $\delta$ must overcome a
threshold value $\delta_c =\frac{\ln(\lambda/\xi)}{\ln(\L/\lambda )}$. 
In the limit $L\to \infty$, due to the logarithmically 
divergent second term in Eq. \ref{nv2}, the magnetic/superconducting 
vortex pair is always energetically favorable. However, in practice 
the logarithmic factor never exceeds $5-7$. Therefore, the value 
$\delta$ exceeding 1 triggers  generation of the MV/SV pairs.
The value of  $\delta$ can be enhanced  by either reducing
the supercarrier density $n_s$ or increasing the thickness of the 
magnetic film.

In a bi-layer of XY ferromagnet and SC films with Curie temperature 
$T_c$ larger than SC transition temperature, the ferromagnetic state 
will be observed in the temperature interval between $T_s$  and $T_c$.  
Below $T_s$  the MV/SV vortices proliferate.  They destroy both, 
ferromagnetism and superconductivity. Hovewer, the reentrant phase transition
at $T=T_V$ leads to restoration of both order parameters if $\delta(T=0)$ is
smaller than the logarithms ratio. 

The vortices of the same sign repulse each other, whereas the 
antivortex attracts vortex.  Together they form a dense system, 
most likewise a crystal at low temperature, with the lattice constant 
$d \sim \lambda$, since $\lambda$  is the only characteristic length 
in the system. This is a new phase, never observed in 2d systems
before. It differs from a vortex crystal in the 3d superconductors in 
magnetic field. It is also very different from the 
Berezinskii-Kosterlitz-Thouless zero-field liquid vortex phases having 
no bound vortex-antivortex pairs.

In conclusion, we presented a general formalism to treat the 
interaction between magnetic textures and superconductors.
As applications, we have shown that vortices in superconducting films 
can be generated by magnetic dots with normal to the film 
magnetization. Superconducting vortices also appear in bound pairs 
with magnetic vortices in a homogeneous, easy-plane magnetic film.
We have found how the magnetic dot size and its material control 
the vortex appearance. Spontaneous generation of the MV/SV pairs
in the bi-layer of the XY magnetic and SC films starts when 
characteristic FM-SC interaction energy $\epsilon_m$ exceeds the 
energy  $\epsilon_0$ necessary for creation of a single SV.   
We have proposed experiments to detect the bound state of the  
magnetic dot and vortex and coupled pairs of magnetic and 
superconducting vortices. 

This work was partly supported by the grants
DE-FG03-96ER45598, NSF DMR-00-72115 and THECB ARP 010366-003. 
It is a pleasure to acknowledge discussions with A.I. Larkin,
D.R. Nelson, D.Naugle, G.W. Crabtree and I.K. Schuller.

\end{multicols}


\begin{references}
\bibitem[**]{byline} In early experiments 
magnetic dots were a few microns in size \cite{spmsi1}.
\end{references}
\end{document}